\documentclass[12pt,english]{article}
\usepackage[english]{babel}
\usepackage[utf8]{inputenc}
\usepackage{natbib,authblk,bbm,bm}
\usepackage{amsmath,setspace,geometry,lineno,amsfonts,dsfont}
\usepackage[vlined]{algorithm2e}
\SetKwInOut{Input}{Input}
\SetKwInOut{Output}{Output\,}
\SetKwInOut{Data}{Data}
\SetKwProg{Tree}{Tree}{}{EndTree}
\usepackage{graphicx}
\usepackage[colorinlistoftodos]{todonotes}
\usepackage{xcolor,colortbl}
\usepackage{fullpage}
\usepackage{mathtools}
\usepackage{framed}
\colorlet{shadecolor}{lightgray!20}
\usepackage{tikz}
\usepackage{bbm}
\usepackage{hyperref}
\usetikzlibrary{fit}
\usetikzlibrary{calc}
\usetikzlibrary{decorations.pathreplacing}
\usetikzlibrary{arrows.meta}
\tikzset{>={Latex[width=1.5mm,length=1.5mm]}}
\title{Penalized estimation of flexible hidden Markov \\ models for time series of counts}

\author{Timo Adam$^{1}\footnote{Corresponding author; email: \texttt{timo.adam@uni-bielefeld.de}.}$, Roland Langrock$^1$, and Christian H.\ Weiß$^2$\\ $^1$Bielefeld University, Germany\\  $^2$Helmut-Schmidt-University Hamburg, Germany}

\begin{document}
\begin{spacing}{1.25}
\maketitle
\end{spacing}
\vspace{-10mm}
\begin{spacing}{1.5}

\begin{abstract}
Hidden Markov models are versatile tools for modeling sequential observations, where it is assumed that a hidden state process selects which of finitely many distributions generates any given observation. Specifically for time series of counts, the Poisson family often provides a natural choice for the state-dependent distributions, though more flexible distributions such as the negative binomial or distributions with a bounded range can also be used. However, in practice, choosing an adequate class of (parametric) distributions is often anything but straightforward, and an inadequate choice can have severe negative consequences on the model's predictive performance, on state classification, and generally on inference related to the system considered. To address this issue, we propose an effectively nonparametric approach to fitting hidden Markov models to time series of counts, where the state-dependent distributions are estimated in a completely data-driven way without the need to select a distributional family. To avoid overfitting, we add a roughness penalty based on higher-order differences between adjacent count probabilities to the likelihood, which is demonstrated to produce smooth probability mass functions of the state-dependent distributions. The feasibility of the suggested approach is assessed in a simulation experiment, and illustrated in two real-data applications, where we model the distribution of i) major earthquake counts and ii) acceleration counts of an oceanic whitetip shark (\textit{Carcharhinus longimanus}) over time.
\end{abstract}


\section{Introduction}
\label{Introduction}

Over the past half century, hidden Markov models (HMMs) have become increasingly popular tools for modeling time series data where, at each point in time, a hidden state process selects among a finite set of possible distributions for the observations \citep{zuc16}. For example, in economic applications, the states of the Markov chain underlying the observations (which determines the state process) are often good proxies for market regimes such as periods of economic growth or recessions, while in ecology, they can regularly be linked to an animal's behavioral modes such as resting, foraging, or traveling. Other areas of application include volcanology \citep{beb07}, psychology \citep{vis02}, and medicine \citep{jac02}, to name but a few examples. Depending on the application at hand, potential aims that can be addressed using HMMs are manifold, including the prediction of future values of a time series, decoding of the hidden states underlying the observations, and inference on the drivers for example on the state-switching dynamics.

HMMs constitute a versatile framework for modeling diverse types of time series data and can easily be tailored for example to binary data \citep{sch12}, positive real-valued data \citep{lan12b}, circular data \citep{bul12}, categorical data \citep{mar12}, compositional data \citep{lan13}, or count data \citep{lag15}. Here we specifically focus on the latter type of data, i.e.\ sequences of non-negative integers. \citet{mac00} provide an introduction to the use of HMMs for discrete-valued time series, including count data, while \citet{weiss18} gives a more comprehensive overview of the various classes of models for time series of counts as well as further types of discrete-valued time series. A typical example for a count time series is the number of corporate defaults observed on a monthly, quarterly, or yearly basis: In periods of economic growth, these may be generated by some distribution with relatively small mean, whereas during recessions, another distribution with relatively higher mean may be active. Although the market regime is not directly observable, it clearly affects the observed counts (c.f.\ \citealp{li15}). For an application that is similar in spirit, see \citet{ham18}, where both the number and the amount of operational losses of a bank are modeled using an HMM-type approach. Beyond economics, HMMs have been applied to time series of counts in a variety of different fields, including, \textit{inter alia}, medicine (multiple sclerosis leisure counts; \citealp{alt05}), geology (volcanic eruption counts; \citealp{beb07}), epidemiology (oliomyelitis counts; \citealp{le99}), ecology (pilot whale vocalization counts; \citealp{pop17}), and bioinformatics (T-lymphocyte counts; \citealp{mar18}). 

Specifically for time series of counts, a Poisson distribution often provides a natural choice for the state-dependent distributions within an HMM, where one rate parameter is estimated for each state of the underlying Markov chain. While more flexible distributions such as the negative binomial can also be used, or models for bounded counts like the binomial, choosing an adequate family of parametric distributions generally remains a difficult task in practice, with potentially severe negative consequences in case an inadequate choice is made. For the case of continuous-valued time series, \cite{lan15} propose a nonparametric approach to estimating the state-dependent distributions within an HMM, where the state-dependent densities are constructed using penalized B-splines (P-splines; \citealp{eil96}), which yields flexible yet smooth functional shapes of the resulting densities without the need to make any distributional assumptions. For time series of counts that are either naturally bounded or considered to be effectively bounded (defining an upper threshold for the support on which the distribution is modeled), one can, in principle, easily avoid a distributional assumption by directly estimating the values of the state-dependent probability mass functions (p.m.f.s) on the support considered; this approach is in fact implemented in the R package \texttt{hmm.discnp} \citep{turner18}. However, without penalization such an approach will often lead to overfitting, which again limits the usefulness of corresponding models in particular for prediction and classification. Following \cite{sco80} and \cite{sim83}, and similar in spirit to \citet{lan15}, we here suggest to address this issue by considering a penalized likelihood function, where we add a roughness penalty based on higher-order differences between adjacent count probabilities. We further demonstrate how the penalty term can be adjusted in presence of (e.g.\ zero-) inflated observations, where small differences between corresponding count probabilities and their respective neighbors are not necessarily desired. This conceptually simple approach is demonstrated to be immensely effective in producing reliable estimates of both simple and complex state-dependent p.m.f.s within an HMM, with a smoothing parameter adjusting the required flexibility in a completely data-driven way.

The paper is structured as follows: In Section \ref{sec2}, we briefly recall the model formulation and dependence structure of HMMs, and introduce some notation specifically for the case of time series of counts. Furthermore, we provide an efficient algorithm for evaluating the likelihood and discuss how the model parameters can be estimated in a penalized maximum likelihood framework. In Section \ref{sec3}, we assess our approach in a simulation experiment, where we compare the performance of the penalized nonparametric approach to its parametric counterpart. In Section \ref{sec4}, we present two real-data applications, where we model the distribution of i) major earthquake counts and ii) acceleration counts of an oceanic whitetip shark (\textit{Carcharhinus longimanus}) over time.

\section{Methodology}
\label{sec2}

\subsection{A nonparametric hidden Markov model for count data}
\label{sec2.1}

Basic HMMs comprise two stochastic processes, only one of which is observed, namely the time series of interest, $\{ Y_t\}_{t=1, \dots, T}$. The observed process is driven by another, hidden process, which we denote by $\{ S_t\}_{t=1, \dots, T}$. The latter process is the so-called state process, which is usually modeled as an $N$-state Markov chain. Throughout this paper we consider a first-order Markov chain, i.e.\ we assume the state process to satisfy the Markov property, $\Pr(S_{t} | S_1, \dots, S_{t-1}) = \Pr(S_{t} | S_{t-1})$, $t=2,\dots,T$. This simplifying dependence assumption is exploited in the likelihood calculations provided in Section \ref{sec2.2} and can be relaxed to higher-order Markov chains or semi-Markov chains if deemed necessary \citep{zuc16}. Assuming the first-order Markov chain to be time-homogeneous, the state transition probabilities are summarized in the $N \times N$ transition probability matrix (t.p.m.) $\boldsymbol{\Gamma}$, with elements
\begin{equation*}
\gamma_{ij} = \Pr \left(S_{t} = j | S_{t-1} = i\right),
\end{equation*}
$i,j = 1, \dots, N$. The initial state probabilities, i.e.\ the probabilities of the state process being in the different states at time $t=1$, are summarized in the $N$-dimensional row vector $\boldsymbol{\delta}$, with elements
\begin{equation*}
\delta_i = \Pr \left(S_1 = i\right),
\end{equation*}
$i = 1, \dots, N$. If the Markov chain is assumed to be stationary, which is reasonable in many applications, then the initial distribution is the stationary distribution, i.e.\ the solution to the equation system $\boldsymbol{\delta} \boldsymbol{\Gamma} = \boldsymbol{\delta}$ subject to $\sum_{i=1}^N \delta_i = 1$ \citep{zuc16}. Otherwise, the initial state probabilities are parameters which need to be estimated. The state process is completely specified by the initial state probabilities and the state transition probabilities.

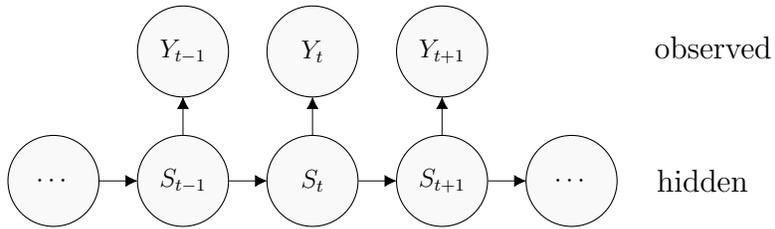
\begin{figure}[t!]
\centering
\begin{tikzpicture}[node distance = 2cm]
\tikzset{state/.style = {circle, draw, minimum size = 42pt, scale = 0.82}}
\node [state,fill=lightgray!10] (6) [] {$S_{t}$};
\node [state,fill=lightgray!10] (5) [left = 5mm of 6] {$S_{t-1}$};
\node [state,fill=lightgray!10] (7) [right = 5mm of 6] {$S_{t+1}$};
\node [state,fill=lightgray!10] (8) [above = 5mm of 5] {$Y_{t-1}$};
\node [state,fill=lightgray!10] (9) [above = 5mm of 6] {$Y_{t}$};
\node [state,fill=lightgray!10] (10) [above = 5mm of 7] {$Y_{t+1}$};
\node [state,fill=lightgray!10] (15) [left = 5mm of 5] {$\cdots$};
\node [state,fill=lightgray!10] (16) [right = 5mm of 7] {$\cdots$};
\node [text=black,] (17) [right = 21mm of 7] {hidden};
\node [text=black,] (18) [above = 12mm of 17] {\hspace*{1.5mm} observed};
\draw[->, black, line width=0.2pt] (15) to (5);
\draw[->, black, line width=0.2pt] (7) to (16);
\draw[->, black, line width=0.2pt] (5) to (6);
\draw[->, black, line width=0.2pt] (6) to (7);
\draw[->, black, line width=0.2pt] (5) to (8);
\draw[->, black, line width=0.2pt] (6) to (9);
\draw[->, black, line width=0.2pt] (7) to (10);
\end{tikzpicture}
\caption{Dependence structure of a basic hidden Markov model.}
\label{fig1}
\end{figure}

The basic dependence structure is such that the observations are assumed to be conditionally independent of each other, given the states. The states then directly select which of $N$ possible distributions generates the observation at any time point. This dependence structure is illustrated in Figure \ref{fig1}. For time series of counts, the observed (state-dependent) process $\{ Y_t\}_{t=1,\dots,T}$ is determined by its state-dependent p.m.f.s, $\Pr(Y_t=k|S_t=i)$, $k=0,\dots,K$ and $i=1,\dots,N$. It is common to use some distributional family such as the class of Poisson distributions to model these state-dependent p.m.f.s (see, e.g., \citealp{alt05,beb07}). Here we do not make any such assumption, and instead assign a state-specific probability to each possible count on the bounded support $\{0, \dots, K\}$, i.e.\ we consider the model parameters
\begin{equation*}
\pi_{i,k} = \Pr(Y_t=k|S_t=i),
\end{equation*}
$i=1,\dots,N$ and $k=0, \dots, K$. While count data can be unbounded, we here consider an upper threshold as to obtain a fixed number of parameters. The support should be bounded in a reasonable way; specifically, it should at least cover all observed counts. The state-dependent process is completely specified by the count probabilities ($N K$ free model parameters). Thus, with this model formulation, we consider a possibly large number of parameters rather than say only one (as e.g.\ in the case of a Poisson or binomial distribution) or two (as e.g.\ in the case of a negative binomial distribution). 

Although the parameter space in this model formulation is still finite-dimensional, it will usually have a fairly high dimension, with the individual parameters $\pi_{i,k}$ not being of direct interest themselves. As a consequence, we follow \citet{turner18} and call our approach ``nonparametric''. In particular, this label emphasizes that the state-dependent p.m.f.s are not determined by a small number of parameters, as would be the case when a distributional family such as the class of Poisson distributions would be used. In addition, with the given model formulation we are not restricted to any particular functional shape of the state-dependent p.m.f.s, and instead have full flexibility to let the data speak for themselves, just like with other methods for which the label ``nonparametric'' is commonly used in the literature.

\subsection{Likelihood evaluation}
\label{sec2.2}

For a given parameter vector $\boldsymbol{\theta}$, in our case comprising the state transition probabilities $\gamma_{ij}$, $i,j=1,\ldots,N$ with $i\neq j$, and the count probabilities $\pi_{i,k}$, $i=1,\dots,N$ and $k=0, \dots, K-1$, and a time series of counts $y_1,\dots,y_T$, the likelihood function of the HMM as formulated above is given by
\begin{equation}
\mathcal{L}(\boldsymbol{\theta}) = \Pr(y_1,\dots,y_T | \boldsymbol{\theta}) = \boldsymbol{\delta} \mathbf{P}(y_1) \prod_{t=2}^T \boldsymbol{\Gamma} \mathbf{P}(y_t) \mathbf{1},
\label{llh}
\end{equation}
where $\mathbf{P}(k) = \text{diag}(\pi_{1,k}, \dots, \pi_{N,k})$, and where $\mathbf{1} \in \mathbb{R}^N$ is a column vector of ones. The evaluation of (\ref{llh}) corresponds to the application of the so-called forward algorithm. Specifically, defining the forward probabilities $\alpha_t(i) = \Pr(y_1, \dots, y_t, S_t = i)$, which are summarized in the row vectors $\boldsymbol{\alpha}_t = \bigl(\alpha_t(1), \dots, \alpha_t(N)\bigr)$, the recursive scheme
\begin{equation}
\begin{split}
\boldsymbol{\alpha}_1 &= \boldsymbol{\delta} \mathbf{P} \left(y_1\right);\\
\boldsymbol{\alpha}_t &= \boldsymbol{\alpha}_{t-1} \boldsymbol{\Gamma} \mathbf{P}\left(y_t\right),
\end{split}
\label{fa}
\end{equation}
$t = 2, \dots, T$, can be applied to arrive at $\boldsymbol{\alpha}_T$, from which we obtain  $\mathcal{L}(\boldsymbol{\theta})= \Pr(y_1,\dots,y_T | \boldsymbol{\theta})$ $= \sum_{i=1}^N \alpha_T(i) = \boldsymbol{\alpha}_T \mathbf{1}$ by the law of total probability (cf.\ \citealp{zuc16}).

Using (\ref{fa}), evaluating the likelihood requires $\mathcal{O}(TN^2)$ operations, which makes an estimation of the model parameters by numerically maximizing the likelihood (or log-likelihood in case of numerical underflow) practically feasible even for relatively long time series and a moderately high number of states. Alternatively, the expectation-maximization algorithm, which also arrives at a (local) maximum of the likelihood, can be used. For the model formulation considered here, the latter approach is implemented in the R package \texttt{hmm.discnp} \citep{turner18}.

\subsection{Roughness penalization}
\label{sec2.3}

The downside of the above nonparametric and hence very flexible approach to estimating the state-dependent p.m.f.s is its propensity to overfit any given data. Especially in cases where $T$ is small relative to the number of model parameters, $N(N-1)+ NK$, the fitted state-dependent p.m.f.s will often be anything but smooth, and may even involve isolated spikes with implausible gaps in between (corresponding to a lack of data in regions where observations would in fact be expected to occur in the long run). For short time series, it can in fact easily happen that for specific values well within the plausible range of observations to occur in future, each state-dependent probability is estimated to be zero, namely if no such observations are present in the training data (see Section \ref{sec4.1} for an example of this problem). The consequence of this would be that the model deems the corresponding values to be impossible to occur in future, which could be problematic for forecasting. To avoid such kind of overfitting, we add a roughness penalty to the logarithm of the likelihood given in (\ref{llh}), which leads to the penalized log-likelihood
\begin{equation}
\log \left(\mathcal{L}_\text{pen.} (\boldsymbol{\theta}) \right) = \log \left( \mathcal{L} (\boldsymbol{\theta}) \right) - \sum_{i=1}^N \lambda_i \sum_{k=m}^K \left(\Delta^m \pi_{i,k} \right)^2,
\label{equ7}
\end{equation}
where $\lambda_i, i=1,\dots, N$, denotes a smoothing parameter associated with the $i$-th state-dependent distribution, and where $\Delta^m \pi_{i,k} = \Delta^{m-1}(\Delta \pi_{i,k})$, $\Delta \pi_{i,k} = \pi_{i,k} - \pi_{i,k-1}$, denotes the $m$-th order differences between adjacent count probabilities.

The inclusion of the penalty term, together with the associated smoothing parameters, allows to control the variance of the otherwise unrestricted and hence highly variable estimation of the state-dependent p.m.f.s. Optimizing the penalized log-likelihood (\ref{equ7}) then amounts to finding a good compromise between goodness of fit, as measured by the logarithm of the likelihood given in (\ref{llh}), and smoothness of the state-dependent p.m.f.s, as measured by the $m$-th order differences. The order of the differences, $m$, should be chosen pragmatically based on the data at hand. With $m=1$, a uniform distribution is obtained as $\lambda_i \rightarrow \infty$ (the penalty term vanishes if all count probabilities are equal). For $m=2$, we obtain a triangular distribution in the limit (the penalty term vanishes if all count probabilities lie on a straight line with arbitrary slope). Based on our experience, $m=3$ or an even higher order produces the most reliable estimates across a range of scenarios of varying complexity. 
In case of zero-inflation, which in practice often occurs when dealing with count data, it may make sense not to penalize differences between the probability mass on 0 and the adjacent count probabilities (i.e.\ those on $1, \dots, m$), as otherwise the penalization will shrink the estimate of $\pi_{i,0}$ and increase its neighboring count probabilities as to ensure smoothness of the resulting state-dependent p.m.f.s, which in case of a genuine excess of zeros can be undesirable (see Section \ref{sec4.2} for an example of this problem). The penalty term in (\ref{equ7}) can then be replaced by the inflation-adjusted penalty term $\sum_{i=1}^N \lambda_i \sum_{k=m+1}^K (\Delta^m \pi_{i,k})^2$, such that the $\pi_{i,0}$'s are estimated without any constraints related to the smoothness of the resulting state-dependent p.m.f.s.

Regarding the choice of $K$, i.e.\ the size of the support on which the conditional p.m.f.s are estimated, it is again necessary to use a $K$ greater than or equal to the highest count observed. However, with unbounded counts, it does in fact make sense to choose a somewhat larger $K$, as the absence of values greater than $K$ in the training data does not guarantee that such values will not occur in future. While without penalization, a nonparametric approach would estimate the probability of such future events to be 0, the penalized approach will place positive probability on counts slightly larger than the maximal value observed due to the enforced smoothing. 

\subsection{Model fitting}
\label{sec2.4}

For a given set of smoothing parameters, maximum penalized likelihood estimates for the model parameters can be obtained by numerically maximizing the penalized log-likelihood (\ref{equ7}) using some Newton-Raphson-type optimization routine, e.g.\ the R function \texttt{nlm} \citep{r17}. To increase the chance of having found the global rather than a local maximum, a multiple-start-point strategy can be applied, where the penalized log-likelihood (\ref{equ7}) is maximized from different, possibly randomly selected starting values, then choosing the estimate corresponding to the highest penalized log-likelihood \citep{zuc16}.

In our model, all parameters to be estimated are probabilities. The corresponding constraints to be satisfied are as follows:
\begin{enumerate}
\item[(1)] All model parameters to be estimated lie in the interval $[0,1]$;
\item[(2)] the rows of the t.p.m.\ sum to one, i.e.\ $\sum_{j=1}^N \gamma_{ij} = 1$, $i=1,\dots,N$;
\item[(3)] the initial state probabilities sum to one, i.e.\ $\sum_{i=1}^N \delta_i = 1$;
\item[(4)] the count probabilities of each state-dependent distribution sum to one, i.e.\ $\sum_{k=0}^K \pi_{i,k}$ $= 1$, $i=1,\dots,N$.
\end{enumerate}
To ensure that these restrictions are satisfied, the constrained model parameters can be transformed into unconstrained ones using multinomial logit links, i.e.\
\begin{equation}
\gamma_{ij} = \frac{\exp(\gamma_{ij}^*)}{\sum_{l=1}^N \exp(\gamma_{il}^*)}, ~ \delta_i = \frac{\exp(\delta_i^*)}{\sum_{l=1}^N \exp(\delta_l^*)}, ~ \text{and} ~ \pi_{i,k} = \frac{\exp(\pi_{i,k}^*)}{\sum_{l=0}^K \exp(\pi_{i,l}^*)},
\label{trans_unconstr}
\end{equation}
then maximizing the penalized log-likelihood (\ref{equ7}) with respect to the unconstrained parameters $\gamma_{ij}^*, \delta_i^*$, and $\pi_{i,k}^*$, fixing $\gamma_{ii}^*=\delta_1^*=\pi_{i,0}^*=0$, $i=1\ldots,N$, for identifiability. The constrained parameters are then obtained by applying the transformation (\ref{trans_unconstr}). Regarding identifiability in general, and specifically in case of the very flexible model formulation considered here, \citet{ale16} show that for the HMM to be identifiable it is sufficient if the t.p.m.\ has full rank and the state-dependent distributions are distinct, conditions that can be expected to be met in most practical settings where HMMs seem natural candidate models.

An adequate choice of the smoothing parameters is crucial for finding a good balance between model fit and estimator variance. We here adopt the cross-validation approach as presented in \cite{lan15}, where we try to find the optimal vector $(\lambda_1^*,\ldots,\lambda_N^*)$ from a pre-specified, $N$-dimensional grid using a greedy search algorithm:
\begin{enumerate}
\item[(1)] Choose an initial vector $(\lambda^{(0)*}_1,\dots,\lambda^{(0)*}_N)$ from the grid and set $r=0$;
\item[(2)] calculate the average out-of-sample log-likelihood for $(\lambda^{(r)*}_1,\dots,\lambda^{(r)*}_N)$ and each direct neighbor on the grid;
\item[(3)] from these values choose the vector $(\lambda^{(r+1)*}_1,\dots,\lambda^{(r+1)*}_N)$ as the one that yielded the highest out-of-sample log-likelihood averaged across folds;
\item[(4)] repeat steps 2.\ and 3.\ until $(\lambda^{(r+1)*}_1,\dots,\lambda^{(r+1)*}_N)= (\lambda^{(r)*}_1,\dots,\lambda^{(r)*}_N)$.
\end{enumerate}
In step (2), the out-of-sample log-likelihood is evaluated as follows: First, for any given smoothing parameter vector, the out-of-sample observations --- e.g.\ 5\% of all data points in each of 20 folds --- are treated as missing data for model training using maximum penalized likelihood estimation, hence replacing the corresponding $\mathbf{P}(y_t)$-matrices in the likelihood function given in (\ref{llh}) by identity matrices. Then the out-of-sample log-likelihood can be calculated in the same way, now treating the in-sample observations as missing data and using the estimated model parameters to calculate the out-of-sample (unpenalized) log-likelihood.

\section{Simulation experiment}
\label{sec3}

\begin{figure}[t!]
\includegraphics[width=0.4875\textwidth]{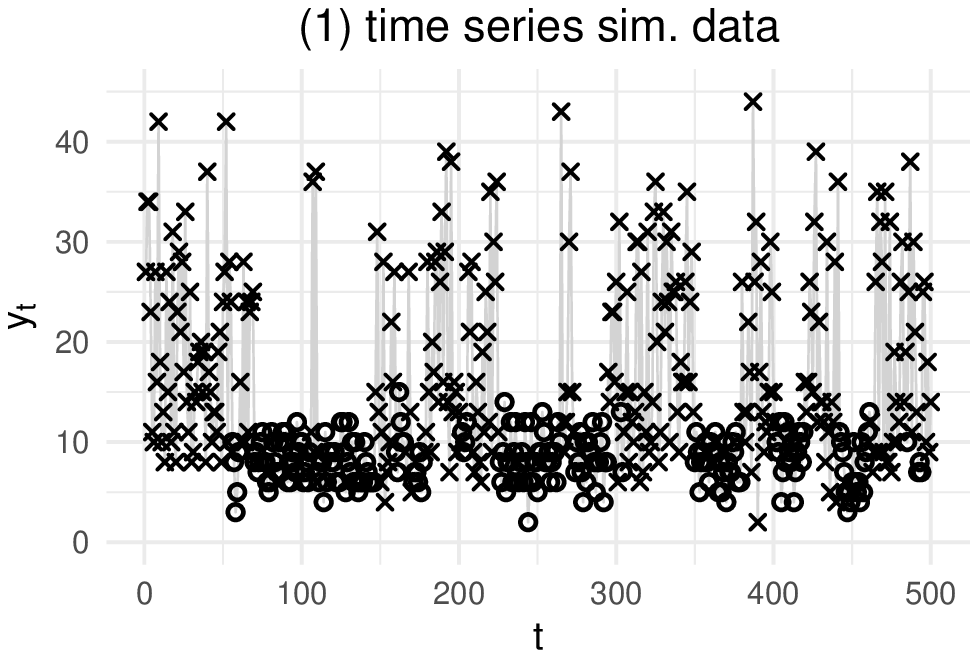}
\includegraphics[width=0.4875\textwidth]{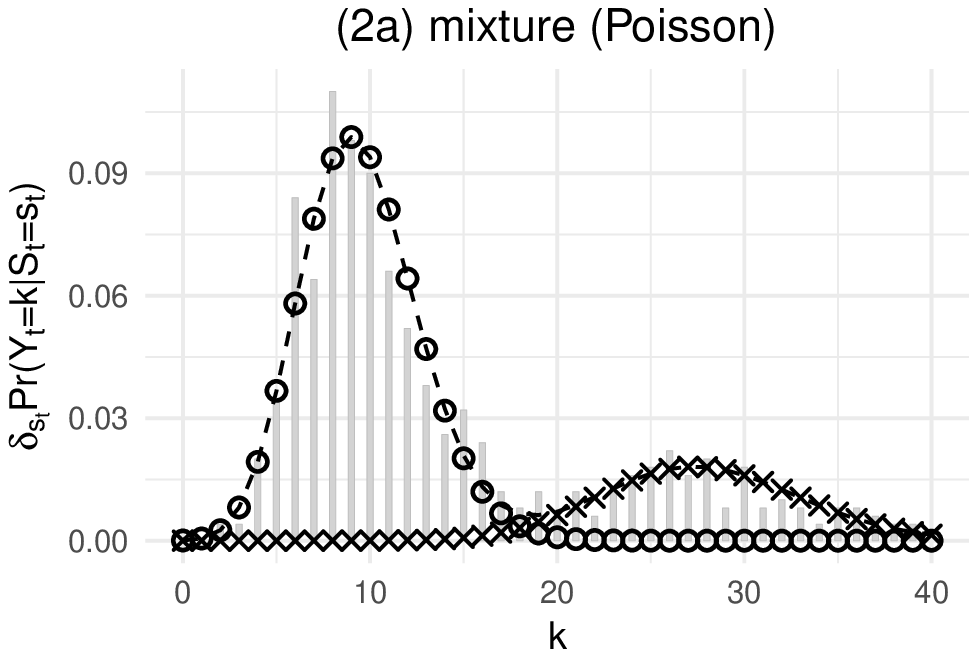}\\
\includegraphics[width=0.4875\textwidth]{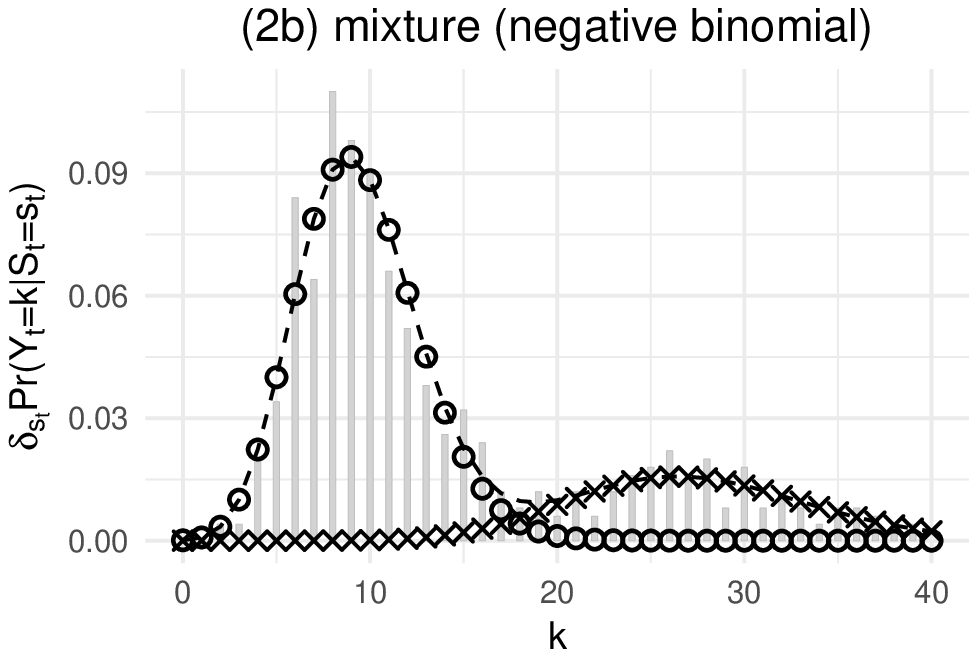}
\includegraphics[width=0.4875\textwidth]{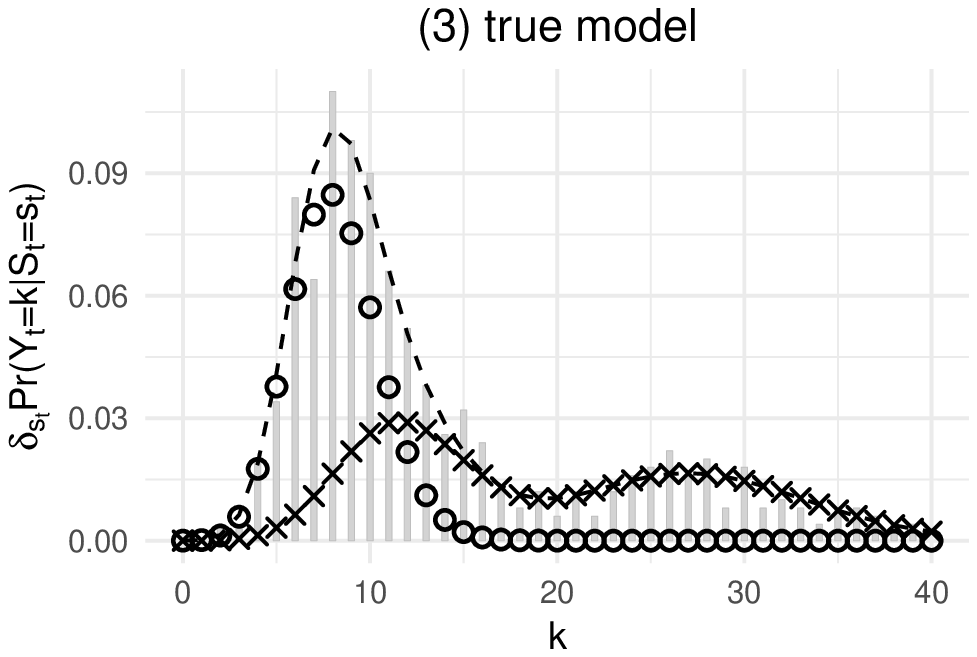}
\caption{Example time series (panel (1)) and bar plots together with the associated fitted parametric mixture models (panels (2a)--(2b)) and the true state-dependent p.m.f.s (panel (3)). Dots and crosses indicate the conditional p.m.f.s; dashed lines the true marginal distribution.}
\label{fig2}
\end{figure}

To demonstrate the feasibility of the suggested approach, and to illustrate its potential usefulness, we conducted the following simulation experiment: In each of $200$ simulation runs, we first simulated $T=500$ realizations from a $2$-state Markov chain with initial (stationary) distribution $\boldsymbol{\delta}=(0.5,0.5)$ and t.p.m.
\begin{equation*}
\boldsymbol{\Gamma} = \begin{pmatrix}
0.95 & 0.05\\
0.05 & 0.95
\end{pmatrix}.
\end{equation*}
Conditional on the simulated states, the observations were then drawn from either of the following distributions: A Conway-Maxwell-Poisson (CMP) distribution (when the state process was in state 1), or a 2-component mixture of a Poisson and a CMP distribution (when in state 2). Compared to the Poisson distribution, the CMP distribution has an additional parameter, which allows to account for under- or overdispersion relative to the Poisson. The shape of the two state-dependent p.m.f.s, and additionally an example time series as simulated in one of the simulation runs, are displayed in panels (1) and (3) in Figure \ref{fig2}. The marginal distribution could be fairly well captured by a mixture of two Poisson or two negative binomial distributions (c.f.\ panels (2a)--(2b) in Figure \ref{fig2}), such that a 2-state Poisson or negative binomial HMM would seem to be a natural choice for modeling. However, the underlying state-dependent distributions do in fact substantially deviate from a Poisson, exhibiting underdispersion in state 1, and overdispersion as well as bimodality in state 2. This fairly complex model formulation was selected as to show the full potential of the new estimation framework, but also to highlight potential pitfalls that may occur when choosing simplistic parametric model formulations based in particular on an inspection of the marginal distribution of the data.

We consider four different model specifications: (1a) The true parametric model, as a benchmark only (noting that, in practice, a model specification as complex as the given one effectively could not be guessed based on an inspection of the marginal distribution of the data; cf.\ Figure \ref{fig2}); (1b) an incorrectly specified parametric model with Poisson state-dependent distributions (which as discussed above would seem to be a reasonable model specification based on the marginal distribution); (2a) the unpenalized nonparametric estimation approach; and (2b) the nonparametric estimation approach with third-order difference penalty. For the latter two, the support of the state-dependent p.m.f.s was taken to be $\max(40,\max(y_1,\dots,y_T))$ in each simulation run. The smoothing parameters required within model (2b) were selected by $20$-fold cross validation over the grid $(10^1,10^2, \dots,10^6) \times (10^1,10^2, \dots,10^6)$. The means of the smoothing parameters selected over the 200 runs were $8173.5$ (state 1) and $570610.0$ (state 2), respectively.

\begin{figure}[t!]
\includegraphics[width=0.4875\textwidth]{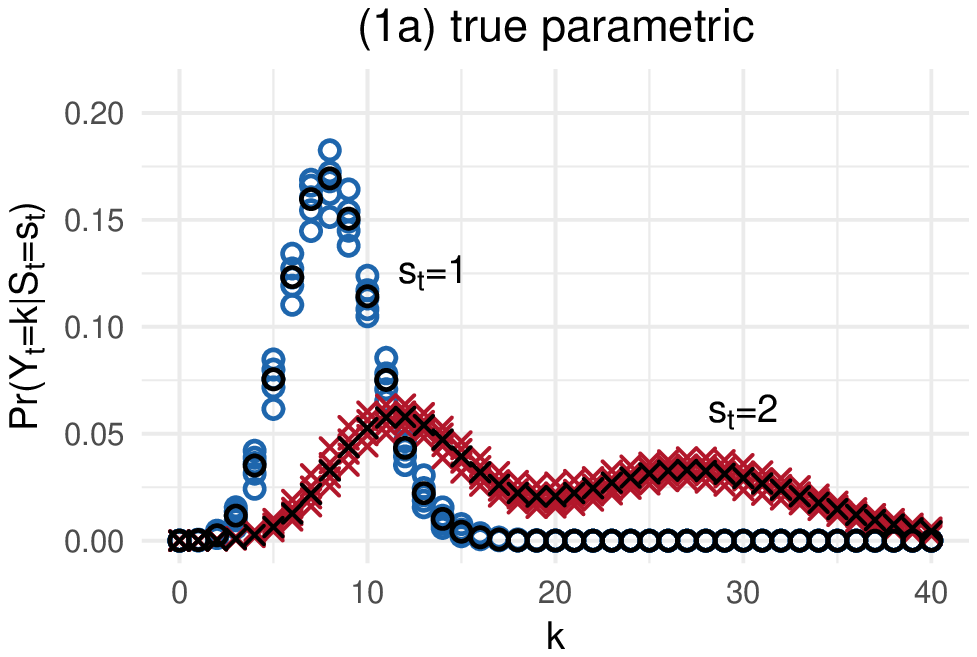}
\includegraphics[width=0.4875\textwidth]{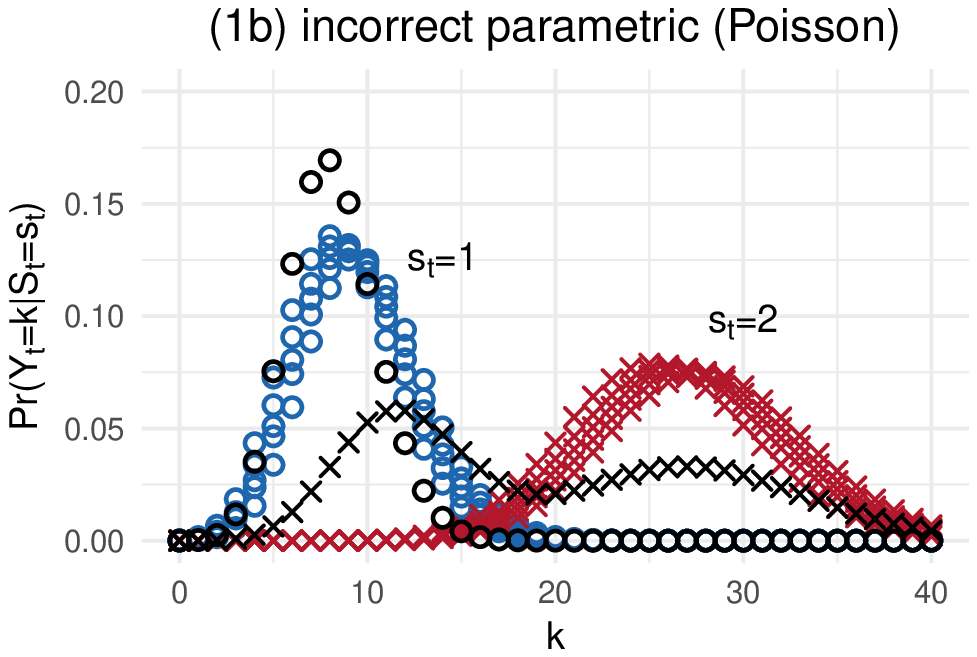}\\
\includegraphics[width=0.4875\textwidth]{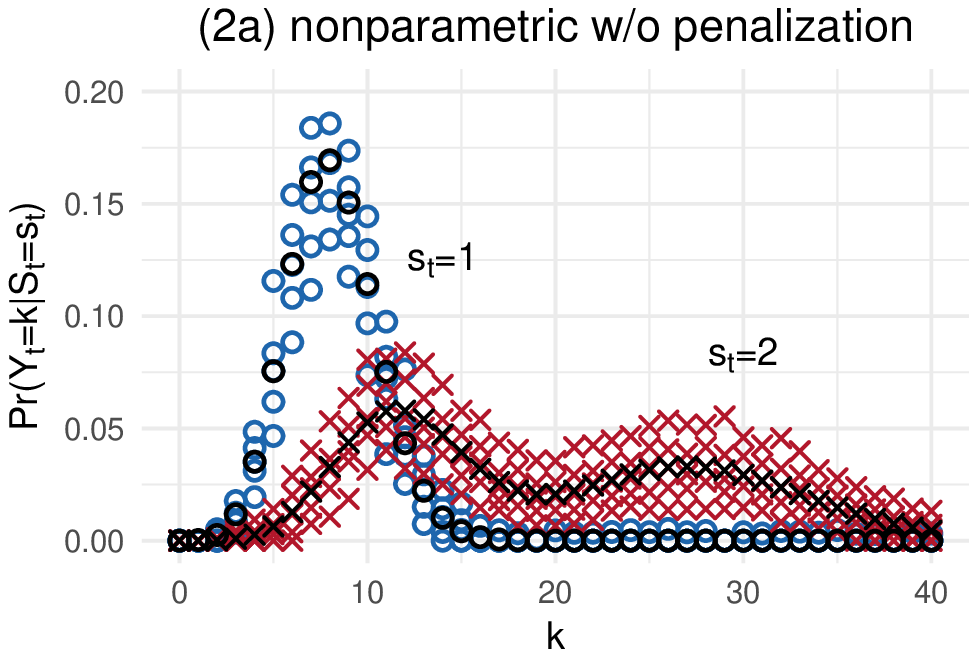}
\includegraphics[width=0.4875\textwidth]{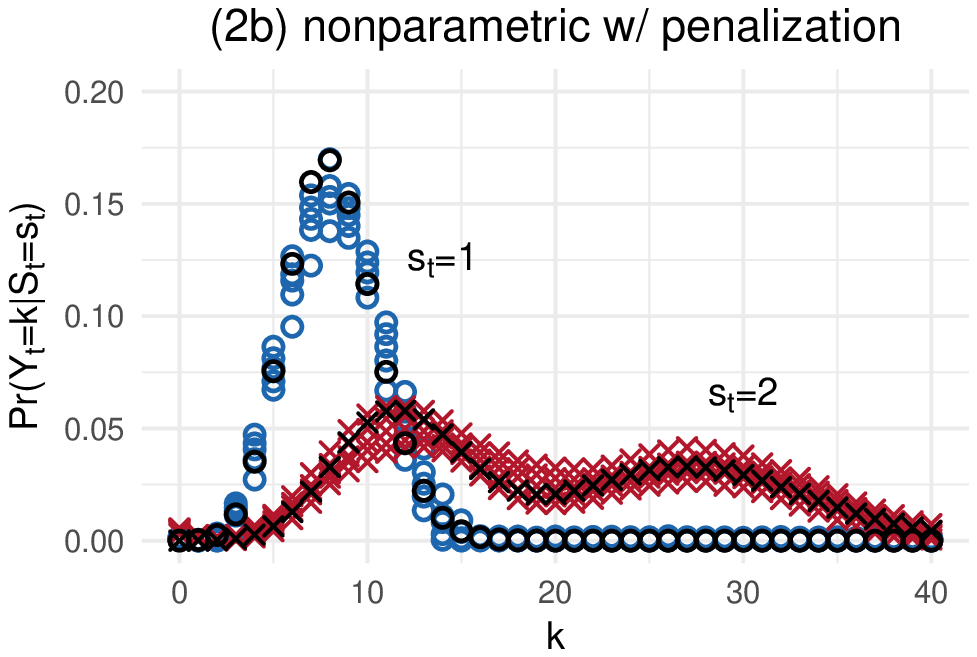}
\caption{Fitted state-dependent p.m.f.s under the different estimation approaches obtained in $200$ simulations runs. The 0.05, 0.25, 0.5, 0.75, and 0.95-quantiles of the estimated count probabilities are indicated by colored dots (state 1) and crosses (state 2); the true state-dependent p.m.f.s are are indicated by black dots and crosses, respectively.}
\label{fig3}
\end{figure}

The fitted state-dependent p.m.f.s under the four different models are illustrated in Figure \ref{fig3}. It can be seen that the proposed penalized nonparametric approach, model (2b), produced estimates very similar to those obtained when using the true parametric model, i.e.\ model (1a) --- estimator variance is slightly higher when using model (2b), and also there is some underestimation of peaks and overestimation of troughs, as would be expected when using this effectively nonparametric estimation approach. In any case, given that the true parametric model specification is unknown in practice, this first impression regarding the performance of the penalized nonparametric approach is encouraging. Regarding the other two competitors, the incorrectly specified parametric model, model (1b), clearly lacks the flexibility to capture the functional shape of the true state-dependent p.m.f.s (thus exhibiting a substantial bias), while the unpenalized nonparametric approach, model (2a), leads to a very much higher estimator variance due to overfitting.

To more formally assess the performance of the suggested approach, we calculated three different performance measures as obtained for each of the four different estimation methods considered: 
\begin{itemize}
\item[(1)] The average Kullback-Leibler divergences (KLDs) between the true and the estimated state-dependent distributions, averaged across all $200$ simulation runs, i.e.
\begin{align*}
\overline{\text{KLD}}_i & = \frac{1}{200} \sum_{r=1}^{200}  \sum_{k=0}^{40} \pi_{i,k} \log \left( \frac{\pi_{i,k}}{\hat \pi_{i,k}^{(r)}} \right),
\end{align*}
where $\hat \pi_{i,k}^{(r)}$ denotes the estimated state-dependent count probabilities in state $i$ as obtained in the $r$-th simulation run;
\item[(2)] the mean absolute errors (MAEs) of the estimated state transition probabilities,
\begin{align*}
\overline{\text{MAE}}_{\hat \gamma_{ij}} & = \frac{1}{200} \sum_{r=1}^{200} \sqrt{ \left(\hat \gamma_{ij}^{(r)} - \gamma_{ij}\right)^2},
\end{align*}
where $\hat \gamma_{ij}^{(r)}$, $i\neq j$, denotes the estimate obtained in the $r$-th simulation run;
\item[(3)] the average state misclassification rates (SMRs),
\begin{equation*}
\overline{\text{SMR}} = \frac{1}{200} \sum_{r=1}^{200} \frac{1}{500} \sum_{t=1}^{500} \mathds{1}_{\hat s_t^{(r)} \neq s_t^{(r)}},
\end{equation*}
where $\hat s_t^{(r)}$ denotes the Viterbi-decoded state at time $t$, and $ s_t^{(r)}$ denotes the true state at time $t$, in the $r$-th simulation run.
\end{itemize}
The different values of these performance measures, as obtained under the four different estimation methods considered, are stated in Table \ref{tab1}. The average time required to fit the penalized nonparametric model (2b) was $9.1$ seconds, which is comparable to the average time required to fit the nonparametric model without penalization (2a) ($6.7$ seconds). The penalized log-likelihood (\ref{equ7}) was evaluated in C++, and maximized using the R function \texttt{nlm} \citep{r17}, on a 3.6 GHz Intel\textregistered{ }Core\texttrademark{ }i7 CPU.

\begin{table}[t!]
\centering
\renewcommand{\arraystretch}{1.25}
\begin{tabular}{m{4.65cm}|m{1.55cm}|m{1.55cm}|m{1.55cm}|m{1.55cm}|m{1.55cm}}
model specification & $\overline{\text{KLD}}_1$ & $\overline{\text{KLD}}_2$  & $\overline{\text{MAE}}_{\hat \gamma_{12}}$ & $ \overline{\text{MAE}}_{\hat \gamma_{21}}$ & $\overline{\text{SMR}}$ \\ \hline
(1a) true parametric & 0.005 & 0.009 & 0.012 & 0.013 & 0.033 \\ \hline
(1b) incorrect parametric & 0.112 & 2.085 & 0.122 & 0.395 & 0.240 \\ \hline
(2a) nonparam.\ w/o pen. & 0.379 & 0.652 & 0.012 & 0.013 & 0.043 \\ \hline
(2b) nonparam.\ w/ pen. & 0.043 & 0.029 & 0.012 & 0.014 & 0.041
\end{tabular}
\caption{Average Kullback-Leibler divergences, mean absolute errors, and average state misclassification rates, as obtained under the different estimation approaches.}
\label{tab1}
\end{table}

Unsurprisingly, the incorrectly specified parametric model (1b) shows the worst performance, with the largest average KLDs (due to the lack of flexibility to capture the shape of the conditional p.m.f.s, in particular within state 2), which notably results in a very high average state misclassification rate ($\overline{\text{SMR}}=0.240$; most observations lower than $15$ are assigned to state $1$, although a considerable number of these observations were actually generated from state $2$) as well as relatively large average MAEs of the estimated transition probabilities ($0.122$ and $0.395$; an obvious consequence of a considerable proportion of observations in $[5,15]$ incorrectly allocated to state 1). The unpenalized nonparametric approach, (2a), as implemented in the R~package \texttt{hmm.discnp} \citep{turner18}, shows a much better performance thanks to its flexibility in particular to capture the bimodality within state 2. Due to the substantial reduction in the variance, the inclusion of the penalty term within the suggested penalized nonparametric approach, (2b), further considerably improves in particular the average deviation of the fitted model from the true model, as measured using the average KLDs. As expected, the true parametric model, (1a), shows the best performance --- in practice, it is however by no means guaranteed that a parametric model formulation close to the true data-generating process can easily be identified.

\section{Real-data case studies}
\label{sec4}

\subsection{Major earthquake counts}
\label{sec4.1}

To illustrate the suggested approach in a real-data application, we model the distribution of major earthquake counts over time. In particular, we consider the observed variable
\begin{equation*}
    y_t=\# \text{ of earthquakes worldwide with magnitude} \geq\text{7 in year }t.
\end{equation*}
The data, which are used as a running example in \cite{zuc16}, cover the period from 1900 to 2006, thus comprising $T=107$ observations in total. As a benchmark for the approach developed in the present paper, we consider a 2-state Poisson HMM as presented in \cite{zuc16}. We additionally fit 2-state HMMs using the unpenalized as well as the novel penalized nonparametric estimation approach. For the latter, we used third-order difference penalties and selected the smoothing parameters via 20-fold cross validation over the grid $(10^1,10^2,\dots,10^{10}) \times (10^1,10^2,\dots,10^{10})$, which led to the optimal values $\lambda_1^*=10^8$ and $\lambda_2^*=10^9$.

\begin{figure}[t!]
\includegraphics[width=0.4875\textwidth]{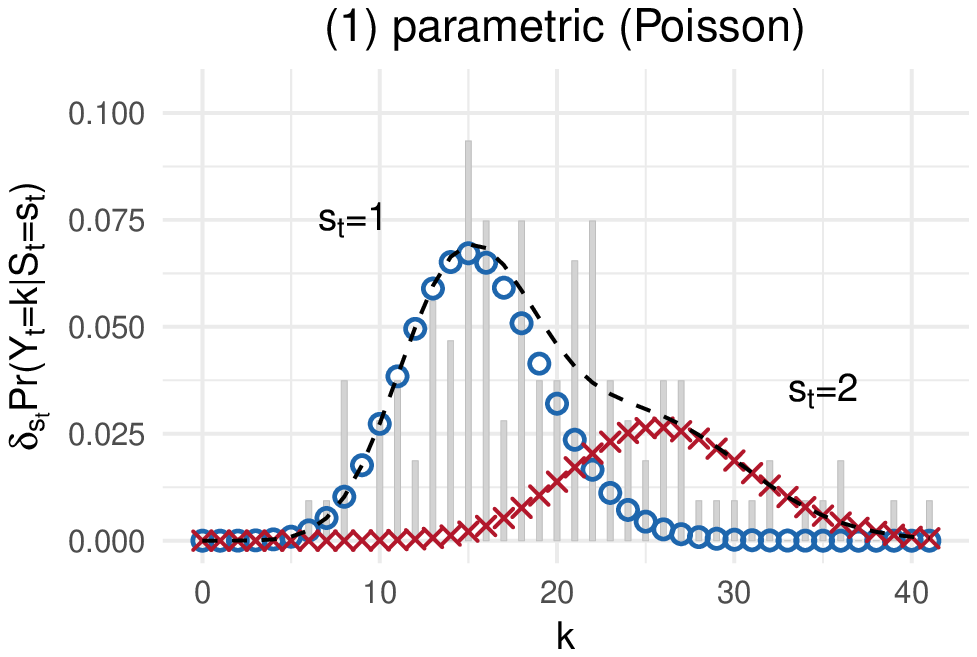}
\includegraphics[width=0.4875\textwidth]{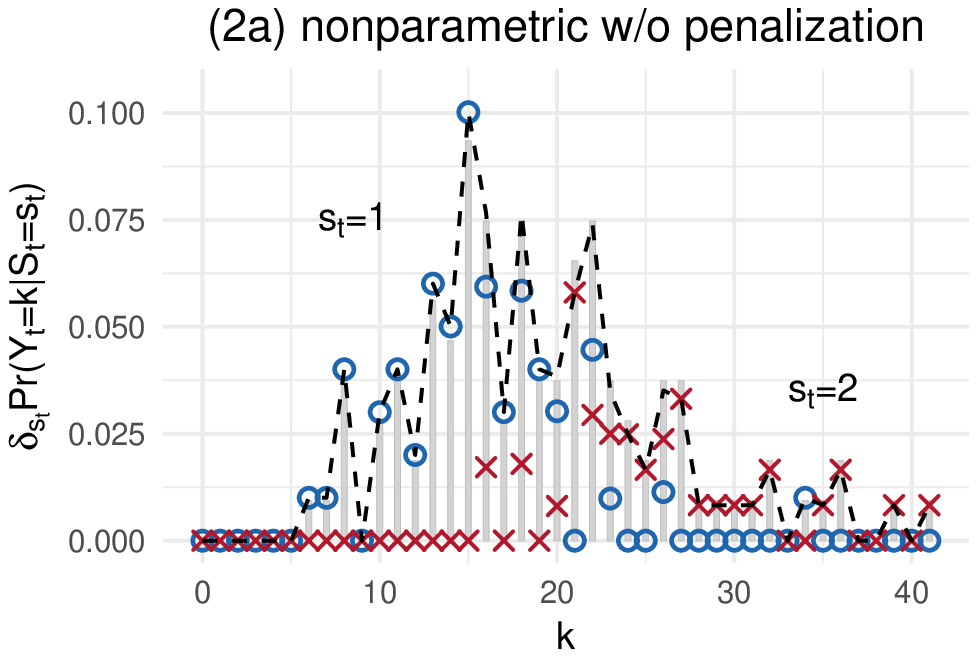}\\
\includegraphics[width=0.4875\textwidth]{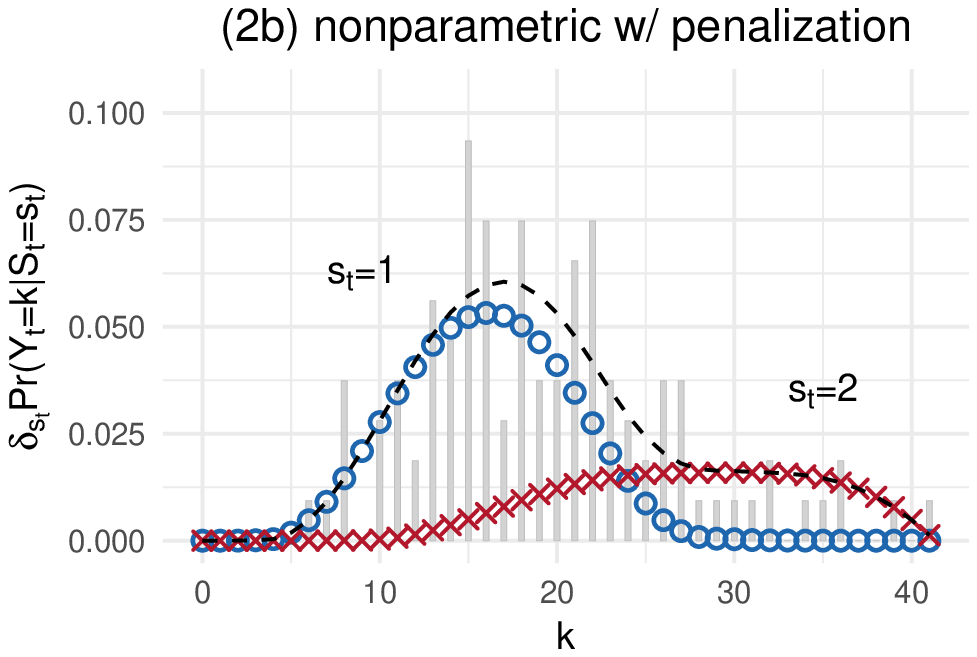}
\includegraphics[width=0.4875\textwidth]{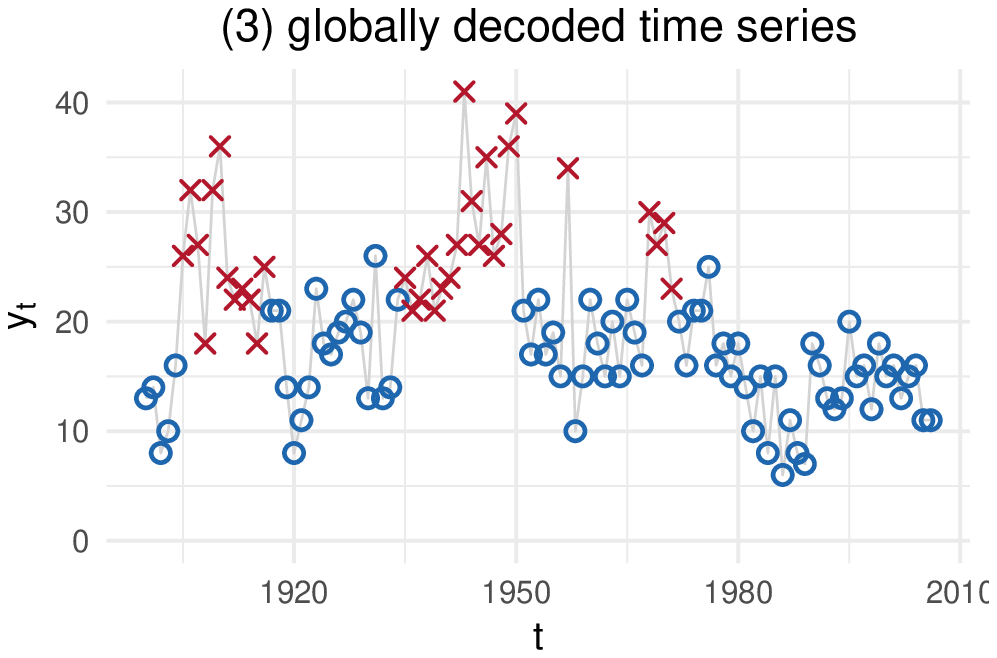}
\caption{Fitted state-dependent p.m.f.s weighted by the stationary state probabilities (colored dots and crosses), together with the associated marginal distributions (black dashed lines), under the different estimation approaches (panels (1)--(2b)). Panel (3) displays the globally decoded time series of major earthquake counts, with the decoding performed under the model fitted using the penalized nonparametric approach.}
\label{fig4}
\end{figure}

With the penalized nonparametric approach, (2b), the t.p.m.\ of the underlying Markov chain was estimated as
\begin{equation*}
\hat{\boldsymbol{\Gamma}} = \begin{pmatrix}
0.934 & 0.066 \\
0.128 & 0.872
\end{pmatrix},
\end{equation*}
such that the associated stationary distribution is $\boldsymbol{\delta}=(0.660, 0.340)$, indicating that about $66$\% and $34$\% of the observations were generated in state $1$ and $2$, respectively. These estimates are very close to those obtained under the 2-state Poisson HMM (cf.\ \citealp{zuc16}). However, as can be seen in Figure \ref{fig4}, the associated state-dependent p.m.f.s as obtained under these two approaches do clearly differ. In particular, the parametric model clearly lacks the flexibility to account for the overdispersion present in the data, particularly within state 2, as captured by the nonparametric approaches. In contrast, due to the short length of the time series, the unpenalized nonparametric approach heavily overfits the data, which demonstrates the need for roughness penalization. In particular, the values 9 and 33 are both assigned a conditional probability of exactly zero in either of the two states --- simply because these counts did not occur between 1900 and 2006 --- such that according to the model these values also cannot occur in future years, which is implausible. Despite the differences in the estimated conditional p.m.f.s, each of the approaches identifies essentially the same pattern, where state 1 may be interpreted as a calm regime with relatively low seismic activity, whereas state 2 corresponds to periods exhibiting relatively high seismic activity. This is further illustrated by means of the globally decoded state sequence under the penalized nonparametric approach, displayed in panel (3) of Figure \ref{fig4}.

\begin{figure}[t!]
\includegraphics[width=0.4875\textwidth]{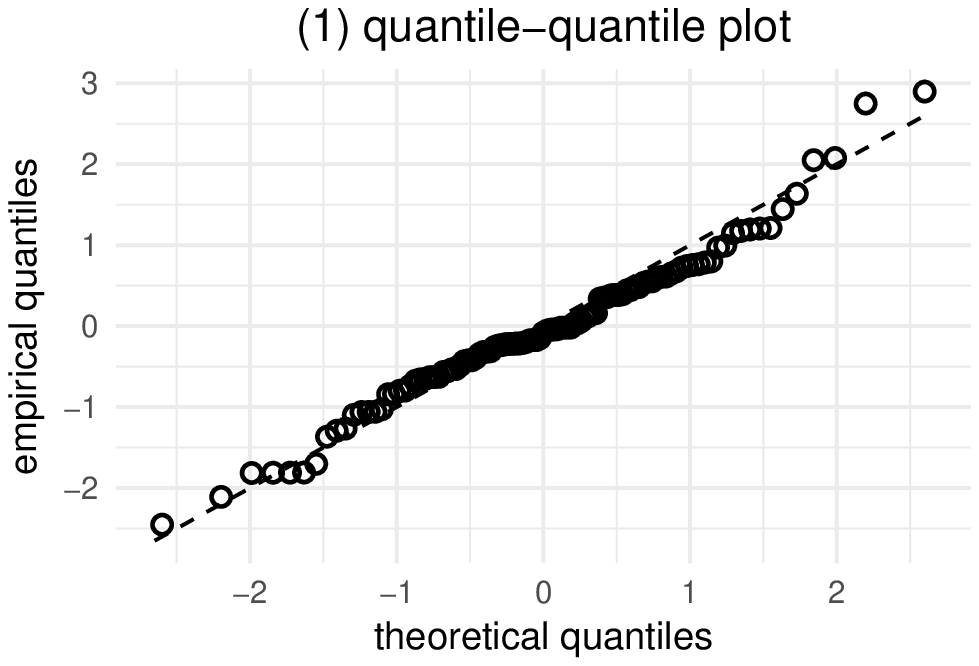}
\includegraphics[width=0.4875\textwidth]{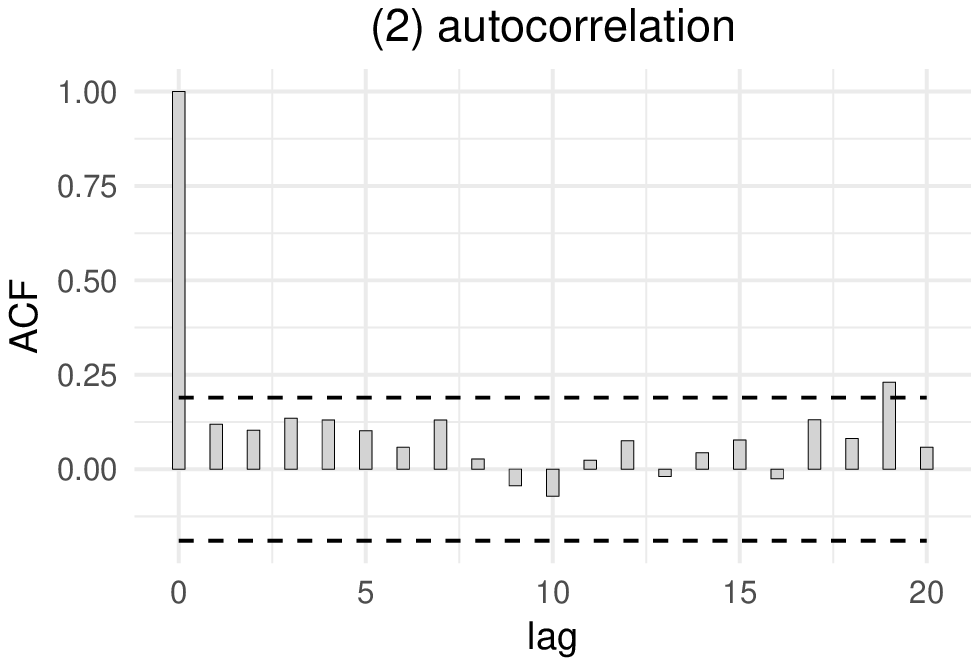}
\caption{Quantile-quantile plot (panel (1)) and autocorrelation function (panel (2)) of the normal ordinary pseudo-residuals as obtained under the penalized nonparametric approach applied to the major earthquake counts.}
\label{fig5}
\end{figure}

For the penalized nonparametric approach, (2b), the quantile-quantile plot as well as the autocorrelation function of the normal ordinary pseudo-residuals are displayed in Figure \ref{fig5}. Ordinary pseudo-residuals indicate whether an observation is extreme relative to the conditional distribution under the fitted model, given all other observations \citep{zuc16}. If the model fits the data well, then the normal ordinary pseudo-residuals approximately follow a standard normal distribution. In that respect, the quantile-quantile plot in Figure \ref{fig5} does not reveal any problematic lack of fit, and the sample autocorrelation function of the pseudo-residuals reveals only little residual autocorrelation. Overall, the model fitted using the penalized nonparametric approach, (2b), shows a satisfactory goodness of fit. In particular, there is no indication that a third state needs to be included in the model, as is the case when using Poisson HMMs (see Chapter 6 in \citealp{zuc16}).

\subsection{Oceanic whitetip shark acceleration counts}
\label{sec4.2}

Oceanic whitetip sharks (\textit{Carcharhinus longimanus}) are circumtropical pelagic apex predators, which are currently listed as ``critically endangered'' in the Northwest Atlantic and ``vulnerable'' globally by the International Union for the Conservation of Nature (IUCN; \citealp{bau06}). Despite an increasing interest in conservation, knowledge about the movement behaviors of this species is very limited \citep{how13}. Here we are particularly interested in activity patterns. From a tri-axial accelerometer, deployed from 8--12 May 2013, we obtained roughly $10.1$ million observations of sway, surge, and heave accelerations at a $30$ Hz resolution. These measures were combined into measures of overall dynamic body acceleration (ODBA) by evaluating the sum over the absolute values of the three different variables, and subsequently averaged over $1$ second intervals (see \cite{leo17} and \cite{lan18} for further background on acceleration data).

In an experimental setting, \cite{lea17} show for another species from the \textit{Carchiarhinus}-family, namely lemon sharks (\textit{Negaprion brevirostris}), that cruising behavior (low activity) corresponds to ODBA values $\leq 0.05$, while bursts in activity correspond to ODBA values $> 0.05$. To reduce the very large size of the data set while maintaining effectively all biologically relevant information on the shark's behavioral dynamics, we thus defined the count variable
\begin{equation*}
    y_t = \sum_{s=1}^{60} \mathds{1}_{\text{ODBA}_{s,t}>0.05},
\end{equation*}
where $\text{ODBA}_{s,t}$ denotes the ODBA value observed during the $s$-th second of the $t$-th minute. The observations $y_t$ --- which are bounded counts with range $\{0,\ldots,60\}$ --- thus simply indicate for how many seconds from any given minute the shark exhibited bursts in activity. The downsampling yielded a total of $T=5560$ observations to be modeled, corresponding to an observation period of about $3.9$ days.

Due to the way the count variable $y_t$ was built, there is an inflation of both 0s and 60s in the data, which within the penalized nonparametric estimation approach, (1c), we account for by not penalizing the differences between the corresponding probabilities and their respective neighbors (i.e.\ 1 \& 2, and 58 \& 59). For comparison purposes, we further analyzed the data using the unpenalized nonparametric approach, (1a), as well as the penalized nonparametric approach without taking the inflation of 0s and 60s into account, (1b). The smoothing parameters for (1c) were selected via 20-fold cross validation over the grid $(10^1,10^2,\dots,10^{10}) \times (10^1,10^2,\dots,10^{10})$, which led to the optimal values $\lambda_1^*=10^4$ and $\lambda_2^*=10^7$ (for comparison, these values were also used for (1b)).

\begin{figure}[t!]
\includegraphics[width=0.4875\textwidth]{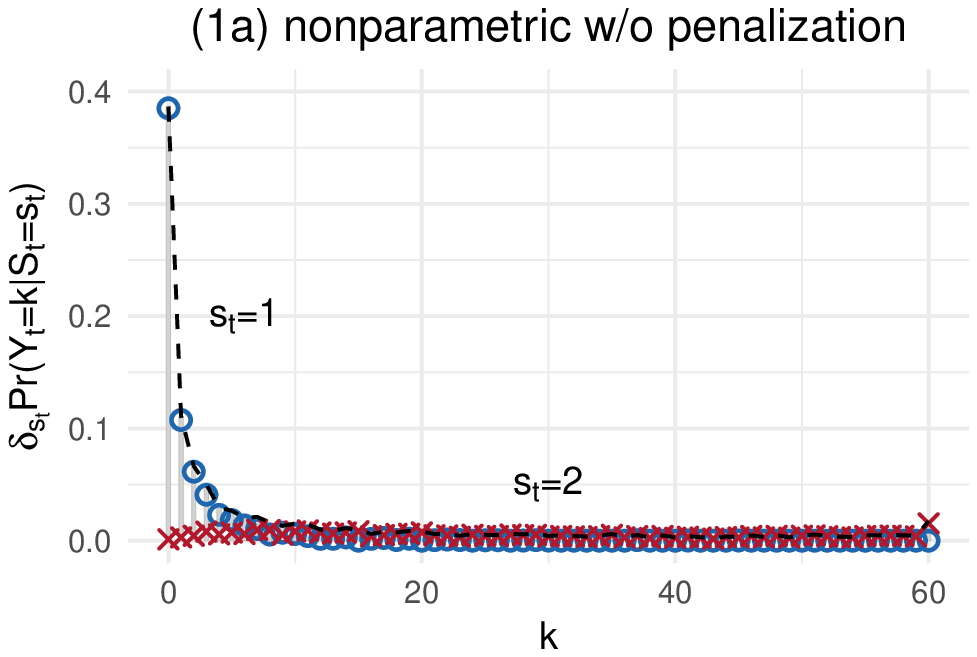}
\includegraphics[width=0.4875\textwidth]{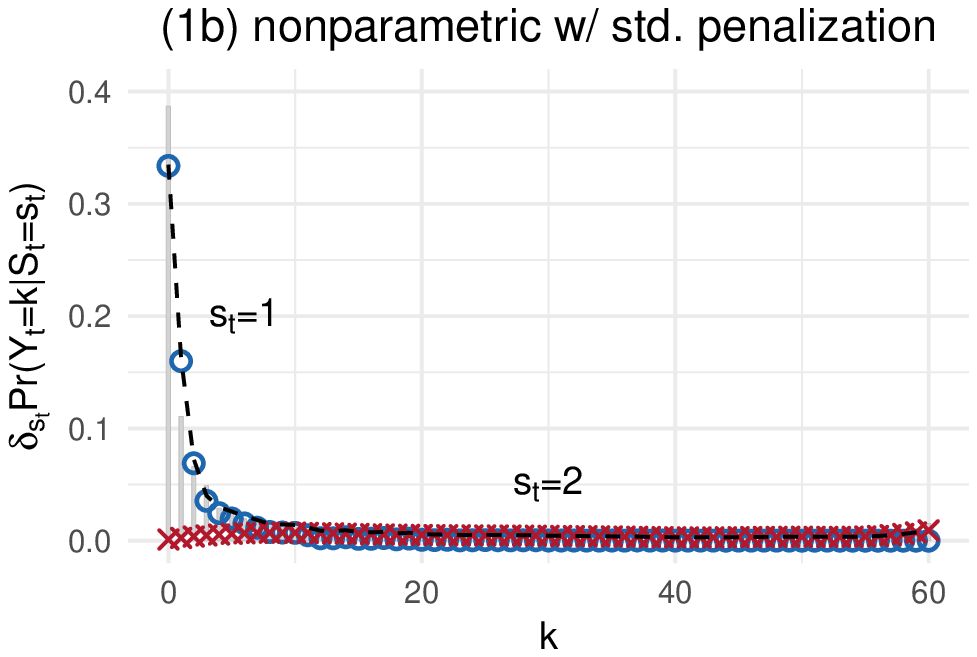}\\
\includegraphics[width=0.4875\textwidth]{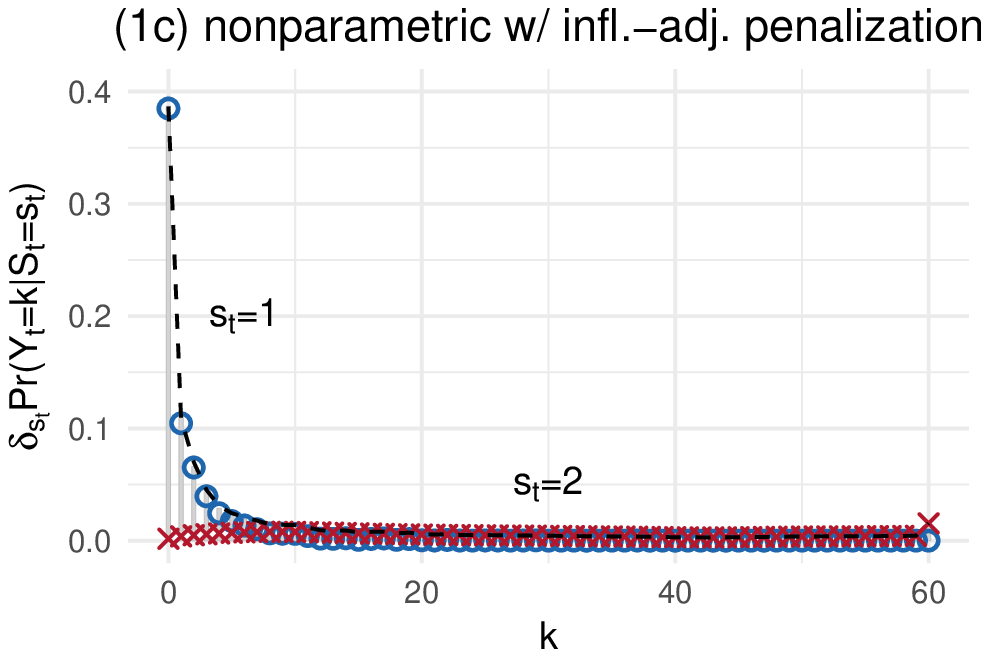}
\includegraphics[width=0.4875\textwidth]{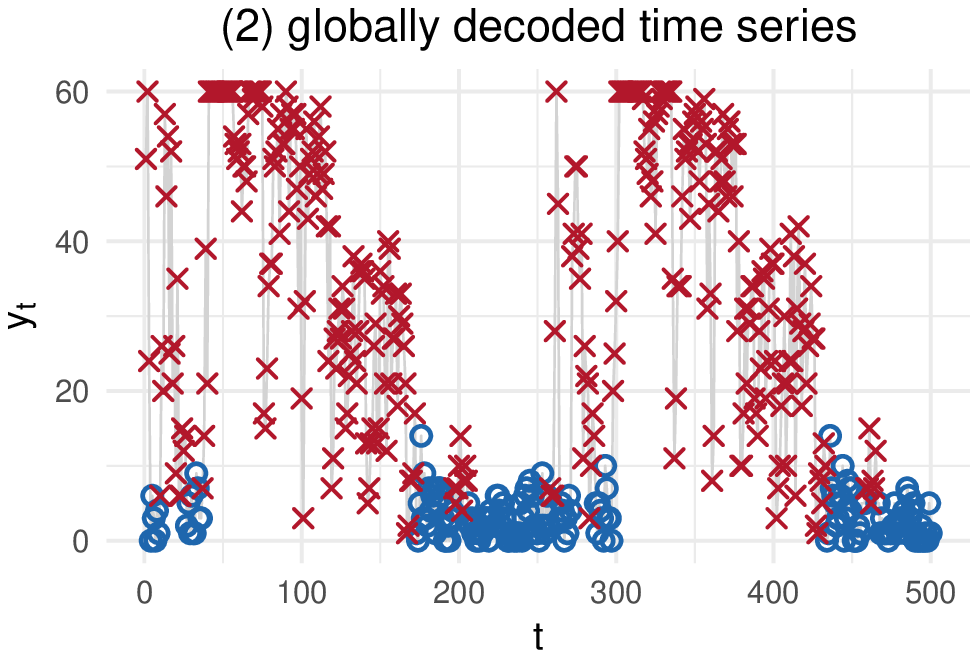}
\caption{Fitted state-dependent p.m.f.s weighted by the stationary state probabilities (colored dots and crosses), together with marginal distributions (black dashed lines), under the different estimation approaches (panels (1a)--(1c)). Panel (2) displays the first 500 observations of the globally decoded time series of acceleration counts, with the decoding performed under the model fitted using the penalized nonparametric approach.}
\label{fig6}
\end{figure}

With the penalized nonparametric approach, (1c), i.e.\ accounting for the inflation of 0s and 60s, the t.p.m.\ of the underlying Markov chain was estimated as
\begin{equation*}
\hat{\boldsymbol{\Gamma}} = \begin{pmatrix}
0.960 & 0.040 \\
0.093 & 0.907
\end{pmatrix},
\end{equation*}
such that the associated stationary distribution is $\boldsymbol{\delta} = (0.697, 0.303)$, indicating that about $70$\% and $30$\% of the observations were generated in state $1$ and $2$, respectively. The state-dependent p.m.f.s fitted with the different estimation approaches, and the globally decoded time series of acceleration counts as obtained under the penalized nonparametric approach, (1c), are displayed in Figure \ref{fig6}.
Since it does not take the excess 0s and 60s into account when penalizing, using the standard penalty term, (1b), results in an underestimation of the probabilities of these counts, and an overestimation of the probabilities of their respective neighbors. Without any penalization, i.e.\ with (1a), as well as with the inflation-adjusted penalty term, i.e.\ with (1c), this bias is avoided. The latter approach yields slightly smoother state-dependent p.m.f.s., but overall the models estimated with (1a) and (1c) are very similar --- this is due to the relatively large sample size, which reduces the importance of penalization. Focusing on the penalized nonparametric approach, (1c), we find that state $1$ captures low activity counts, with a high probability mass on 0 ($\hat \pi_{1,0} = 0.546$), which may therefore be interpreted as a resting/cruising state. State $2$ relates to moderate and high activity counts, with a (relatively) high probability mass on 60 ($\hat \pi_{2,60} = 0.052$), which may therefore be linked to a traveling/hunting behavior.

\begin{figure}[t!]
\includegraphics[width=0.4875\textwidth]{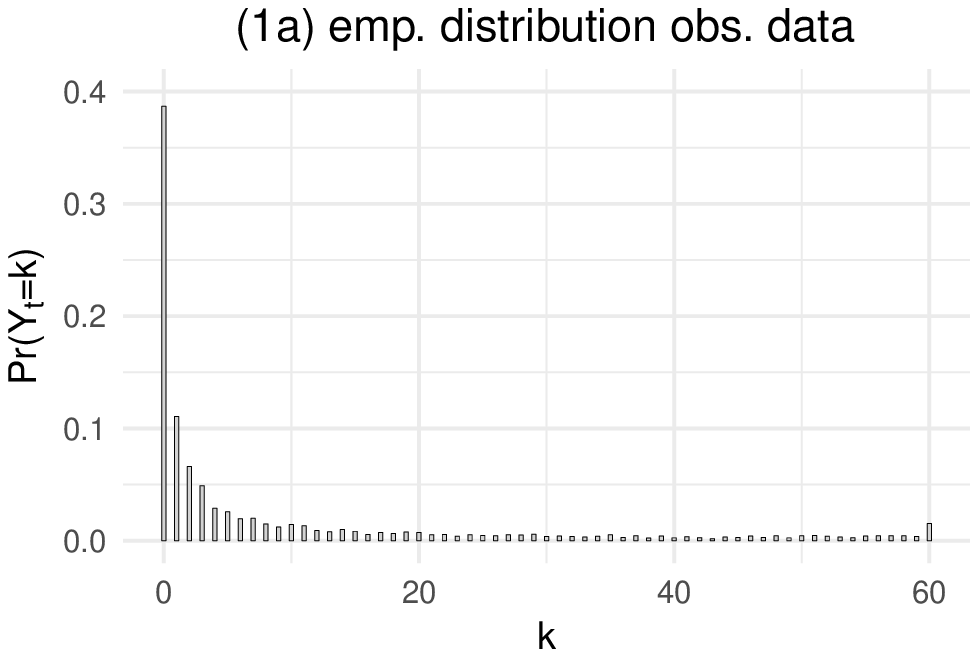}
\includegraphics[width=0.4875\textwidth]{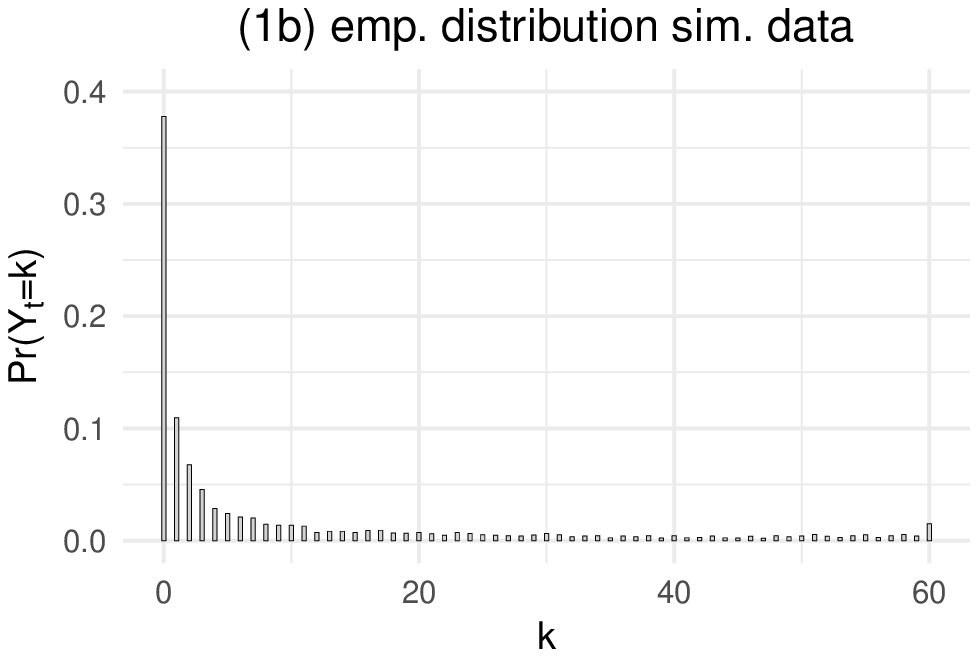}\\
\includegraphics[width=0.4875\textwidth]{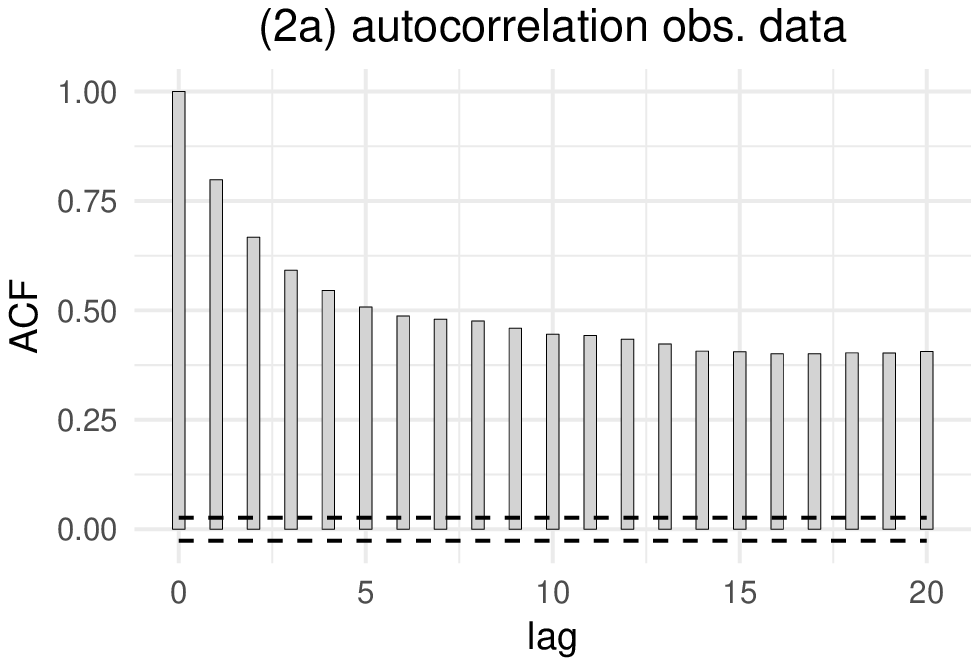}
\includegraphics[width=0.4875\textwidth]{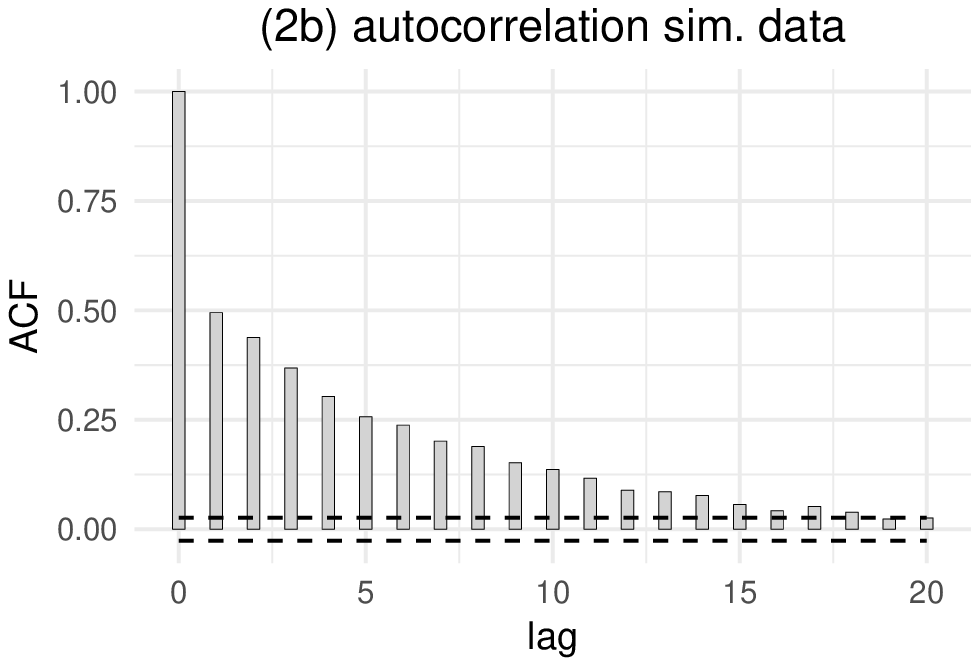}
\caption{Empirical distributions (panels (1a)--(1b)), and autocorrelation functions (panels (2a)--(2b)) of the observed data and the data simulated from the model fitted using the penalized nonparametric approach.}
\label{fig7}
\end{figure}

In terms of model checking, pseudo-residuals are not very helpful in this example due to the excess of zeros in the count data (the quantile-quantile plot, which is not shown here, is dominated by these). Instead, we simply simulate data from the model fitted using the penalized nonparametric approach, (1c), and compare the patterns found in the simulated data, in particular regarding the marginal distribution and the correlation structure, to those found in the empirical data (see Figure \ref{fig7}). Though slightly less formal than a residual analysis, such a simple check, which is similar in spirit to a posterior predictive check as commonly done in Bayesian analyses, is often very helpful in revealing any potential lack of fit; see also Section 2.4 in \citet{weiss18}. Indeed, we find that our model captures the marginal distribution of the data very well (cf.\ panels (1a)--(1b) in Figure \ref{fig7}) --- which is no surprise given its nonparametric nature --- but fails to replicate the strong serial correlation present in the real data (cf.\ panels (2a)--(2b) in Figure \ref{fig7}). In particular, the sequence of the first 500 acceleration counts, as shown in panel (2) of Figure \ref{fig6}, clearly indicates that our assumption of the observations being conditionally independent, given the states, is violated: activity levels seem to sometimes gradually decrease or increase over time, a pattern that is not accounted for within our model formulation. Furthermore panel (2) in Figure \ref{fig6} indicates periodicity in the shark's behavioral patterns, which could potentially be further investigated by incorporating corresponding covariates (e.g.\ the time of day) into the state process, or alternatively using hidden semi-Markov models, which would allow us to more explicitly model the duration of a stay in any given state \citep{lan12}. However, given that this example is supposed to only illustrate the new methodology, we here refrain from considering more complex model formulations that could help to address these nonstandard patterns. 

\section{Discussion}
\label{sec5}

We introduced a penalized nonparametric estimation framework for flexibly tailoring an HMM to a time series of counts. The proposed methodology was shown to be a powerful alternative to parametric HMMs, and may be superior when simple parametric state-dependent distributions are not able to capture some of the features in the data. Specifically, the increased flexibility to capture complex distributional shapes can improve the accuracy of time series forecasts, can reduce state misclassification rates, and can help to avoid making biased inference related, in particular, to the dynamics of the state process. In any case, the approach can also be regarded as an exploratory tool, which can be applied when it is unclear which distributional family should be used for the state-dependent distributions. However, a simple parametric family is to be preferred whenever appropriate, as it will usually be easier to implement and to interpret, and also the computational effort will be much lower than when using the penalized nonparametric approach.

A notorious difficulty with HMMs, which can partly be addressed using the suggested penalized nonparametric approach, is the choice of the number of states. When using model selection criteria to choose an adequate number of states, then these tend to point to models with more states than can plausibly be interpreted, especially (though not only) in ecological applications \citep{poh17}. Inflexibility of (parametric) state-dependent distributions to capture particular features in the data is a common cause of this problem --- such inflexibility can always be compensated for by including additional states that will then usually have no meaningful interpretation anymore \citep{lan15}. In this respect, the effectively unlimited flexibility of the nonparametric approach can help to reduce the required number of states, which will often substantially improve interpretability. In fact, with the new approach only a single state is required to capture the marginal distribution, and potential additional states only need to be included if they help to capture the dependence structure. On the other hand, model selection using information criteria is in fact more difficult within the nonparametric estimation framework, as it is necessary to somehow derive the effective number of parameters used to fit the model \citep{lan18}, a challenge we did not address in the present paper.

In our simulation experiment and real-data case studies, the size of the support on which the state-dependent p.m.f.s were to be modeled was moderate --- the largest support was $\{0,1,\ldots,60\}$ (in the acceleration data example). If p.m.f.s on much larger supports, say on $\{1,\ldots,5000\}$, are to be modeled, then the high dimensionality of the parameter space can become problematic, especially with regard to the computing time. In those instances, one option would be to simply treat the data as stemming from a continuous distribution, building the state-dependent distributions as linear combinations of B-splines \citep{lan15}. Alternatively, the data could be binned, say in intervals $I_1=\{1,\ldots,10\}$, $I_2=\{11,\ldots,20\}$, $\ldots$, $I_{500}=\{4991,\ldots,5000\}$, then estimating conditional p.m.f.s defined on the intervals instead of directly on the counts (in the given example thus reducing the size of the support by a factor 10). In fact, these considerations highlight that the approach considered is by no means restricted to time series of counts. Instead, \textit{any} time series data where observations are at least of ordinal scale can, in principle, be modeled using the approach developed in this work (in the same spirit as presented in \citealp{sim83}). This we believe could be relevant in particular for data on Likert-type scales. Finally, our penalization approach could potentially also be adapted to different types of models for count time series; as an example, one may think of combining it with the semiparametric estimation approach for integer-valued autoregressive (INAR) models by \citet{drost09}. 

\section*{Acknowledgements}

The authors wish to thank Yannis Papastamatiou and Yuuki Watanabe for providing the oceanic whitetip shark data.

\renewcommand\refname{References}
\makeatletter
\renewcommand\@biblabel[1]{}

\end{spacing}


\markboth{}{}
\begin{thebibliography}{9}
\markboth{}{}

\bibitem[\protect\citeauthoryear{Alexandrovich {\em et~al.\/}}{2016}]{ale16}
Alexandrovich, G., Holzmann, H.\ \& Leister, A.\ (2016):
Nonparametric identification and maximum likelihood estimation for hidden Markov models.
\textit{Biometrika}, {\textbf{103}}, 423--434.

\bibitem[\protect\citeauthoryear{Altman \& Petkau}{2005}]{alt05}
Altman, R.M.\ \& Petkau, A.J.\ (2005):
Application of hidden Markov models to multiple sclerosis lesion count data.
\textit{Statistics in Medicine}, \textbf{24} (5), 2335--2344.

\bibitem[\protect\citeauthoryear{Baum {\em et~al.\/}}{2006}]{bau06}
Baum, J., Medina, E., Musick, J.A.\ \& Smale, M.\ (2006):
Carcharhinus longimanus. \textit{IUCN 2012 Red list of threatened species}. URL \url{www.iucnredlist.org}. Accessed 2 July 2012.

\bibitem[\protect\citeauthoryear{Bebbington}{2007}]{beb07}
Bebbington, M.S.\ (2007):
Identifying volcanic regimes using hidden Markov models.
\textit{Geophysical Journal International}, {\textbf{171}} (2),  921--942.

\bibitem[\protect\citeauthoryear{Bulla {\em et~al.\/}}{2012}]{bul12}
Bulla, J., Lagona, F., Maruotti, A.\ \& Picone, M.\ (2012):
A multivariate hidden Markov model for the identification of sea regimes from incomplete skewed and circular time series.
\textit{Journal of Agricultural, Biological, and Environmental Statistics}, {\textbf{17}} (4), 544--567.

\bibitem[\protect\citeauthoryear{Drost {\em et~al.\/}}{2009}]{drost09}
Drost, F.C., van den Akker, R.\ \& Werker, B.J.M. (2009):
Efficient estimation of auto-regression parameters and innovation distributions for semiparametric integer-valued AR(p) models.
\textit{Journal of the Royal Statistical Society: Series~B (Statistical Methodology)}, \textbf{71} (2), 467--485.

\bibitem[\protect\citeauthoryear{Eilers \& Marx}{1996}]{eil96}
Eilers, P.H.C.\ \& Marx, B.D.\ (1996):
Flexible smoothing with B-splines and penalties.
\textit{Statistical Science}, {\textbf{11}}, 89--121.

\bibitem[\protect\citeauthoryear{Hambuckers {\em et~al.\/}}{2018}]{ham18}
Hambuckers, J., Kneib, T., Langrock, R.\ \& Silbersdorff, A.\ (2018):
A Markov-switching generalized additive model for compound Poisson processes, with applications to operational loss models.
\textit{Quantitative Finance}, \textbf{18} (10), 1--20.

\bibitem[\protect\citeauthoryear{Howey-Jordan {\em et~al.\/}}{2013}]{how13}
Howey-Jordan L.A., Brooks, E.J., Abercrombie, D.L., Jordan L.K.B., Brooks, A., Williams, S., Gospodarczyk, E.\ \& Chapman, D.D.\ (2013):
Complex movements, philopatry and expanded depth range of a severely threatened Pelagic Shark, the Oceanic Whitetip (\textit{Carcharhinus longimanus}) in the Western North Atlantic.
\textit{PLOS ONE}, \textbf{8} (2), e56588.

\bibitem[\protect\citeauthoryear{Jackson \& Sharples}{2002}]{jac02}
Jackson, C.H.\ \& Sharples, L.D.\ (2002): 
Hidden Markov models for the onset and progression of bronchiolitis obliterans syndrome in lung transplant recipients.
\textit{Statistics in Medicine}, \textbf{21} (1), 113--128.

\bibitem[\protect\citeauthoryear{Lagona {\em et~al.\/}}{2015}]{lag15}
Lagona, F., Maruotti, A.\ \& Padovano, F.\ (2015):
Multilevel multivariate modelling of legislative count data, with a hidden Markov chain.
\textit{Journal of the Royal Statistical Society: Series A (Statistics in Society)}, \textbf{178} (3), 705--723.

\bibitem[\protect\citeauthoryear{Langrock}{2012}]{lan12b} 
Langrock, R.\ (2012): 
Flexible latent-state modelling of Old Faithful's eruption inter-arrival times in 2009.
\textit{Australian and New Zealand Journal of Statistics}, \textbf{54} (3), 261--279.

\bibitem[\protect\citeauthoryear{Langrock \& Zucchini}{2012}]{lan12}
Langrock, R.\ \& Zucchini, W.\ (2012):
Hidden Markov models with arbitrary state dwell-time distributions.
\textit{Computational Statistics and Data Analysis}, \textbf{55} (1), 715--724.

\bibitem[\protect\citeauthoryear{Langrock {\em et~al.\/}}{2013}]{lan18}
Langrock, R., Adam, T., Leos-Barajas, V., Mews, S., Miller, D.L.\ \& Papastamatiou, Y.P.\ (2018):
Spline-based nonparametric inference in general state-switching models.
\textit{Statistica Neerlandica}, \textbf{72} (3), 179--200.

\bibitem[\protect\citeauthoryear{Langrock {\em et~al.\/}}{2015}]{lan15}
Langrock, R., Kneib, T., Sohn, A.\ \& DeRuiter, S.L.\ (2015):
Nonparametric inference in hidden Markov models using P-splines.
\textit{Biometrics}, \textbf{71}, 520--528.

\bibitem[\protect\citeauthoryear{Langrock {\em et~al.\/}}{2013}]{lan13}
Langrock, R., Swihart, B.J., Caffo, B.S., Crainiceanu, C.M.\ \& Punjabi, N.M.\ (2013):
Combining hidden Markov models for comparing the dynamics of multiple sleep electroencephalograms.
\textit{Statistics in Medicine}, \textbf{32} (19), 3342--3356.

\bibitem[\protect\citeauthoryear{Le Strat \& Carrat}{1999}]{le99}
Le Strat, Y.\ \& Carrat, F.\ (1999):
Monitoring epidemiologic surveillance data using hidden Markov models.
\textit{Statistics in Medicine}, \textbf{18} (24), 3463--3478.

\bibitem[\protect\citeauthoryear{Lear {\em et~al.\/}}{2017}]{lea17}
Lear, K.O., Whitney, N.M., Brewster, L.R., Morris, J.M., Hueter, R.E.\ \& Gleiss, A.C.\ (2017):
Correlations of metabolic rate and body acceleration in three
species of coastal sharks under contrasting temperature regimes.
\textit{Journal of Experimental Biology}, {\textbf{220}}, 397--407.

\bibitem[\protect\citeauthoryear{Leos-Barajas {\em et~al.\/}}{2017}]{leo17}
Leos-Barajas, V., Photopoulou, T., Langrock, R., Patterson, T.A., Watanabe, Y.Y., Murgatroyd, M.\ \& Papastamatiou, Y.P.\ (2017):
Analysis of animal accelerometer data using hidden Markov models.
\textit{Methods in Ecology and Evolution}, \textbf{8} (2), 161--173.

\bibitem[\protect\citeauthoryear{Li \& Cheng}{2015}]{li15}
Li, L.\ \& Cheng, J.\ (2015):
Modeling and forecasting corporate default counts using hidden Markov model.
\textit{Journal of Economics, Business and Management}, \textbf{3} (5), 493--497.

\bibitem[\protect\citeauthoryear{MacDonald \& Zucchini}{2000}]{mac00}
MacDonald, I.L.\ \& Zucchini, W.\ (1997):
Hidden Markov models and other models for discrete-valued time series. Chapman \& Hall/CRC, Boca Raton, FL.

\bibitem[\protect\citeauthoryear{Marino {\em et~al.\/}}{2018}]{mar18}
Marino, M.F., Tzavidis, N.\ \& Alfò, M.\ (2018):
 Mixed hidden Markov quantile regression models for longitudinal data with possibly incomplete sequences.
\textit{Statistical Methods in Medical Research}, \textbf{27} (7), 2231--2246.

\bibitem[\protect\citeauthoryear{Maruotti \& Rocci}{2012}]{mar12}
Maruotti, A.\ \& Rocci, R.\ (2012):
A mixed non‐homogeneous hidden Markov model for categorical data, with application to alcohol consumption.
\textit{Statistics in Medicine}, \textbf{31} (9), 871--886.

\bibitem[\protect\citeauthoryear{Pohle {\em et~al.\/}}{2017}]{poh17}
Pohle, J., Langrock, R., van Beest, F.M.\ \& Schmidt, N.M.\ (2017):
Selecting the number of states in hidden Markov models --- pragmatic solutions illustrated using animal movement.
\textit{Journal of Agricultural, Biological and Environmental Statistics}, \textbf{22} (3), 270--293.

\bibitem[\protect\citeauthoryear{Popov {\em et~al.\/}}{2017}]{pop17}
Popov, V., Langrock, R., DeRuiter, S.L.\ \& Visser, F.\ (2017):
An analysis of pilot whale vocalization activity using hidden Markov models.
\textit{Journal of the Acoustical Society of America}, \textbf{141} (1), 159--171.

\bibitem[\protect\citeauthoryear{R Core Team}{2017}]{r17}
R Core Team (2017),
R: A language and environment for statistical computing. R Foundation for Statistical Computing, Vienna, Austria.
URL \url{https://www.R-project.org}.

\bibitem[\protect\citeauthoryear{Schliehe-Diecks {\em et~al.\/}}{2012}]{sch12}
Schliehe-Diecks, S., Kappeler, P.M.\ \& Langrock, R.\ (2012):
On the application of mixed hidden Markov models to multiple behavioural time series.
\textit{Interface Focus}, \textbf{2} (2), 180--189.

\bibitem[\protect\citeauthoryear{Scott {\em et~al.\/}}{1980}]{sco80}
Scott, D.W., Tapia, R.A.\ \& Thompson, J.R.\ (1980):
Nonparametric probability density estimation by discrete maximum penalized-likelihood criteria.
\textit{The Annals of Statistics}, \textbf{8} (4), 820--832.

\bibitem[\protect\citeauthoryear{Simonoff}{1983}]{sim83}
Simonoff, J.S.\ (1983):
A penalty function approach to smoothing large sparse contingency tables.
\textit{The Annals of Statistics}, \textbf{11} (1), 208--218.

\bibitem[\protect\citeauthoryear{Turner}{2018}]{turner18}
Turner, R.\ (2018):
hmm.discnp: Hidden Markov models with discrete non-parametric observation distributions. 
Version 2.1-5, 2018-11-26.
URL \url{https://cran.r-project.org/package=hmm.discnp}.

\bibitem[\protect\citeauthoryear{Visser {\em et~al.\/}}{2002}]{vis02}
Visser, I., Raijmakers, M.E.J.\ \& Molenaar, P.\ (2002):
Fitting hidden Markov models to psychological data.
\textit{Scientific Programming}, \textbf{10} (3), 185--199.

\bibitem[\protect\citeauthoryear{Wei\ss{}}{2018}]{weiss18}
Wei\ss, C.H.\ (2018):
An introduction to discrete-valued time series.
John Wiley \& Sons, Inc., Chichester.

\bibitem[\protect\citeauthoryear{Zucchini {\em et~al.\/}}{2016}]{zuc16}
Zucchini, W., MacDonald, I.L.\ \& Langrock, R.\ (2016):
Hidden Markov models for time series: An introduction using R, 2nd Edition. Chapman \& Hall/CRC, Boca Raton.

\end{thebibliography}
\end{document}